\documentclass[twoside,onecolumn]{article}%
\usepackage{amssymb}
\usepackage{amsfonts}
\usepackage{amsmath}
\usepackage{graphicx}
\usepackage{xcolor}%
\providecommand{\U}[1]{\protect\rule{.1in}{.1in}}
\newtheorem{theorem}{Theorem}

\newenvironment{proof}[1][Proof]{\noindent\textbf{#1.} }{\ \rule{0.5em}{0.5em}}
\begin{document}

\title{\textbf{Drifting solutions with elliptic symmetry for the compressible
Navier-Stokes equations with density-dependent viscosity }}
\author{Hongli An\thanks{Corresponding Authors and E-mail Addresses:
kaixinguoan@163.com}\\\textit{{\large {College of Science, Nanjing Agricultural University, Nanjing,
210095, PRC}}}\\[0.1in] Manwai Yuen\thanks{E-mail Addresses: nevetsyuen@hotmail.com}\\[-0.07in] \textit{{\large {Department of Mathematics and Information
Technology},}}\\[-0.07in] \textit{{\large {The Hong Kong Institute of Education, 10 Po Ling
Road, Tai Po,}}}\\[-0.07in] \textit{{\large {New Territories, Hong Kong}}}}
\date{Revised 9-May-2014}
\maketitle

\ \newline\textbf{{\large {Abstract:}}} In this paper, we investigate the
analytical solutions of the compressible Navier-Stokes equations with
dependent-density viscosity. By using the characteristic method, we
successfully obtain a class of drifting solutions with elliptic symmetry for
the Navier-Stokes model wherein the velocity components are governed by a
generalized Emden dynamical system. In particular, when the viscosity
variables are taken the same as Yuen in [Yuen M.W. (2008), \textit{Analytical
Solutions to the Navier-Stokes Equations}, J. Math. Phys. \textbf{49},
113102], our solutions constitute a generalization of that obtained by Yuen.


\section{Introduction}

In this paper, we consider the following compressible Navier-Stokes equations
with density-dependent viscosity coefficients
\begin{align}
{\normalsize \rho}_{t}{\normalsize +\mathrm{div}(\rho\mathbf{U})}  &
{\normalsize =}0,\quad\quad\quad\quad\quad\quad\quad\quad\quad\quad\quad
\quad\quad\quad\quad\quad\quad\quad\quad\quad\quad\quad\quad\label{e1}\\
(\rho\mathbf{U})_{t}+\mathrm{{div(\rho\mathbf{U}\otimes\mathbf{U})}%
}-\mathrm{div}(h(\rho)D(\mathbf{U}))-\nabla(g(\rho)\mathrm{div\mathbf{U}%
})+\nabla P(\rho)  &  =\mathbf{0}, \label{e2}%
\end{align}
where $t\in(0,+\infty)$ is the time and $\mathbf{x}\in R^{N}(N\geq2)$ is the
spacial coordinate, while $\rho(x,t)$ denotes the fluid density,
$\mathbf{U}=\mathbf{U}(\mathbf{x},t)=(u_{1},u_{2},\cdots,u_{N})$ stands for
the fluid velocity and $P(\rho)=\kappa\rho^{\gamma}$ for the pressure,
respectively. And
\begin{equation}
D(\mathbf{U})=\frac{\nabla\mathbf{U}+^{t}\nabla\mathbf{U}}{2}%
\end{equation}
is the strain tensor, $h(\rho)$ and $g(\rho)$ are the Lam$\acute{e}$ viscosity
coefficients satisfying
\begin{equation}
h(\rho)>0,\quad\quad\quad h(\rho)+Ng(\rho)\geq0.
\end{equation}

Due to the significance of the Navier-Stokes (NS) equations in various
physical fields such as fluid, plasmas, astrophysics, oceanography and
atmospheric dynamics, the NS equations have been studied extensively and
intensively, which is manifested by a large number of related papers. For
example, the mathematical derivations were derived in the simulation of flow
surface in shallow region \cite{Bresch03,Bresch06, Gerbeau01}. The existence
and uniqueness of the local strong solution were analyzed by Choe and Kim
\cite{HCHK}. While, the existence of global weak solutions was discussed by
Lions \cite{Lions} and other authors \cite{EFAN,YTong,DHHJ,SJPZ,WSSJ}. There
are also some interesting work done on analytical solutions of the NS
equations. For instance, Yuen derived a class of self-similar solutions with
radial symmetry for the NS equations with $h(\rho)=\kappa_{1}\rho^{\theta}$
and $g(\rho)=0$ in \cite{YuenJMP08}. Subsequently, Yuen constructed some
self-similar solutions with elliptic symmetry for the NS equations with
$h(\rho)=\mu$ and $g(\rho)=0$ in \cite{YuenCNSNS}. It is noticed that these
two works were based on the separation method. Recently, by using the
characteristic method, An and Yuen obtained a new class of perturbational
solutions with elliptic symmetry for the NS equations in \cite{AN}.

What needs to point out is that most works mentioned above only hold for NS
equations with special viscosity coefficients $h(\rho)$ and $g(\rho)$. A
natural idea is that the analytical solutions should also exist for the NS
equations with general and reasonable viscosity coefficients $h(\rho)$ and
$g(\rho)$. Since the choice of viscosity coefficients is key to obtain some
physically important solutions. However, up to now, except the work of Guo and
Xin \cite{GuoX}, not much related work has been done. It is remarkable that
here we derive the drifting solutions with elliptic symmetry for the
compressible NS equations with density-dependent viscosity via a
characteristic approach. Interestingly, numerical simulation results show that
such solutions can be used to explain the drifting phenomena of the
propagation wave like Tsunamis in oceans.

\section{Drifting solutions of the NS equations}

Here, we consider the general viscosity coefficients $h(\rho)$ and $g(\rho)$,
which take a form of
\begin{equation}
h(\rho)=\kappa_{1}\rho^{\theta},\quad\quad\quad g(\rho)=\kappa_{2}\rho
^{\theta}. \label{h1}%
\end{equation}
Then, the compressible Navier-Stokes system with density-dependent viscosity
coefficients become
\begin{align}
{\normalsize \rho}_{t}{\normalsize +\mathrm{div}(\rho\mathbf{U})=}0,\quad
\quad\quad\quad\quad\quad\quad\quad\quad\quad\quad\quad\quad\quad\quad
\quad\quad\quad\quad\quad\quad\label{h2}\\
\rho\left[  \mathbf{U}_{t}+\mathbf{U}\cdot\nabla\mathbf{U}\right]
-\mathrm{div}(\kappa_{1}\rho^{\theta}D(\mathbf{U}))-\nabla(\kappa_{2}%
\rho^{\theta}\mathrm{div\mathbf{U}})+\nabla P(\rho)=\mathbf{0}. \label{h3}%
\end{align}
For simplicity, we shall take $D(\mathbf{U})= \nabla\mathbf{U}$ as what has
been chosen by Guo and Xin in \cite{GuoX}.

In the following, we shall give a lemma that proves important to the
constructions of drifting solutions of NS equations with dependent-density
viscosity. \newline\textbf{Lemma} \emph{For the continuity equation of the NS
system, namely:
\begin{equation}
\rho_{t}+\mathrm{div}\left(  \rho\mathbf{U}\right)  =0,\label{e3}%
\end{equation}
there exist solutions
\begin{equation}
\left\{
\begin{array}
[c]{l}%
\rho=\frac{f\left(  \frac{x_{1}-d_{1}}{a_{1}},\frac{x_{2}-d_{2}}{a_{2}}%
,\cdots,\frac{x_{N}-d_{N}}{a_{N}}\right)  }{\underset{i=1}{\overset{N}{\Pi}%
}a_{i}}\\[-0.08in]%
\\
u_{i}=\frac{\dot{a}_{i}}{a_{i}}\left(  x_{i}-d_{i}\right)  +\dot{d}_{i},\text{
\ \ \ \ \ for }i=1,2,\cdots,N
\end{array}
\right. \label{e4}%
\end{equation}
where $d_{i}=d_{i}(t)$, $a_{i}=a_{i}(t)>0$ and an arbitrary $C^{1}$ function
$f\geq0$.}

\begin{proof}
Inspired by the work of \cite{YuenPLA,CRAN}, we perturb the velocity as this
form:
\begin{equation}
\rho=\rho(t,\mathbf{x}),\quad\quad\quad\quad\ u_{i}=\frac{\dot{a}_{i}}{a_{i}%
}\left(  x_{i}-d_{i}\right)  +\dot{d}_{i}. \label{e5}%
\end{equation}
Substitution of this ansatz into the continuity equation (\ref{e3}), yields
\begin{align}
\rho_{t}+\mathrm{div}\left(  \rho\mathbf{U}\right)   &  =\rho_{t}+\nabla
\rho\cdot\mathbf{U}+\rho\nabla\cdot\mathbf{U}\nonumber\\
&  =\frac{\partial}{\partial t}\rho+\underset{i=1}{\overset{N}{\sum}}%
\frac{\partial}{\partial x_{i}}\rho\left[  \frac{\dot{a}_{i}}{a_{i}}%
(x_{i}-d_{i})+\dot{d}_{i}\right]  +\underset{i=1}{\overset{N}{\sum}}\frac
{\rho\dot{a}_{i}}{a_{i}}=0. \label{e6}%
\end{align}
According to the classical characteristic approach \cite{Logan}, we have
\begin{equation}
\frac{dt}{1}=\frac{dx}{\frac{\dot{a}_{i}}{a_{i}}(x_{i}-d_{i})+\dot{d}_{i}%
}=\frac{d\rho}{-\underset{i=1}{\overset{N}{\sum}}\frac{\rho\dot{a}_{i}}{a_{i}%
}} \label{e7}%
\end{equation}
whence, the solution is
\begin{equation}
F\left(  \underset{i=1}{\overset{N}{\Pi}}a_{i}\rho,\frac{x_{1}-d_{1}}{a_{1}%
},\frac{x_{2}-d_{2}}{a_{2}},\cdots,\frac{x_{N}-d_{N}}{a_{N}}\right)  =0
\label{e8}%
\end{equation}
with an arbitrary $C^{1}$ function $F$ such that $\rho\geq0$.

For convenience, we rewrite (\ref{e8}) into an explicit form
\begin{equation}
\rho=\frac{f\left(  \frac{x_{1}-d_{1}}{a_{1}},\frac{x_{2}-d_{2}}{a_{2}}%
,\cdots,\frac{x_{N}-d_{N}}{a_{N}}\right)  }{\underset{i=1}{\overset{N}{\Pi}%
}a_{i}}. \label{e9}%
\end{equation}
Therefore, the proof is completed.
\end{proof}

\textbf{Remark 1:} It is necessary to point out that the negative symbol in
the perturbational\textit{ }\textbf{non-constant }functions $d_{i}$ for the
velocity in (\ref{e4}) is critical to guarantee the use of the characteristic method.

On the application of the above lemma, we construct a class of drifting
solutions with elliptic symmetry for the Navier-Stokes equations
(\ref{h2})-(\ref{h3}). The main result is described as follows:

\begin{theorem}
\label{thm1}For the compressible Navier-Stokes equations with
dependent-density viscosity coefficients, there exists a class of drifting
solutions:
\begin{equation}
\left\{
\begin{array}
[c]{l}%
\rho=\frac{f(s)}{\underset{k=1}{\overset{N}{\Pi}}a_{k}}\\[-0.2in]%
\\
u_{i}=\frac{\dot{a}_{i}}{a_{i}}\left(  x_{i}-d_{i}\right)  +\dot{d}_{i}\text{,
\ \ \ \ \ for }i=1,2,\cdots,N
\end{array}
\right.  \label{e10}%
\end{equation}
where
\begin{equation}
f(s)=\left\{
\begin{array}
[c]{l}%
\alpha e^{-\frac{\xi}{2\theta}s}\text{\quad\quad\quad\quad\quad\quad\quad
\quad\quad\quad\quad\ for }\theta=1\\[0.02in]%
\max\left(  \left(  -\frac{\xi(\theta-1)}{2\theta}s+\alpha\right)  ^{\frac
{1}{\theta-1}},\text{ }0\right)  \text{ \ \ for }\theta\neq1
\end{array}
\right.  \label{e11}%
\end{equation}
with
\begin{equation}
d_{i}=d_{i0}+td_{i1},\quad\quad\quad s=\underset{k=1}{\overset{N}{\sum}%
}\left(  \frac{x_{k}-d_{k}}{a_{k}}\right)  ^{2}.
\end{equation}
In the above $\xi,d_{i0},d_{i1}$ and $\alpha\geq0$ are arbitrary constants.
While the auxiliary functions $a_{i}=a_{i}(t)$ are governed by the generalized
Emden dynamical system:
\begin{equation}
\left\{
\begin{array}
[c]{l}%
\ddot{a}_{i}(t)=\frac{-\xi\left[  k_{1}\sum_{k=1}^{N}\frac{\dot{a}_{k}%
(t)}{a_{k}(t)}+k_{2}\frac{\dot{a}_{i}(t)}{a_{i}(t)}-\kappa\right]  }%
{a_{i}(t)\left(  \underset{k=1}{\overset{N}{\Pi}}a_{k}(t)\right)  ^{\theta-1}%
},\text{ for }i=1,2,\cdots,N\\
a_{i}(0)=a_{i0}>0,\text{ }\dot{a}_{i}(0)=a_{i1}%
\end{array}
\right.  \label{emdengeneral2Chaos}%
\end{equation}
where $a_{i0}$ and $a_{i1}$ are initial conditions. \newline In particular,
for $\xi<0$,\newline(1) if all $a_{i1}<0$, the solutions (\ref{e10}) blow up
on or before the finite time
\begin{equation}
T=\min(-a_{i0}/a_{i1}:a_{i1}<0,\text{ }i=1,2,\cdots,N);
\end{equation}
(2) if all $a_{i1}\geq0$ the solutions (\ref{e10}) exist globally.
\end{theorem}

\textbf{Remark 2:} We emphasize that the intrusion of the viscosity
coefficients $h(\rho)$ and $g(\rho)$ not only makes the solutions are quite
different from what were discussed by Guo et al \cite{GuoX} and Yuen
\cite{YuenJMP08}, but also makes the occurrence of a generalized Emden
dynamical system.

\textbf{Remark 3:} It is known that the Navier-Stokes equations can be used to
describe the drifting phenomena of the propagation wave like Tsunamis in
oceans. Interestingly, numerical simulations fully exhibit such drifting
behaviors. Therefore, we call (\ref{e10}) the drifting solution and the linear
time-dependent functions $d_{i}$ are the drifting terms. When these functions
$d_{i}$ degenerate to constants, namely $d_{i1}=0$ and $d_{i0}=\mathrm{const}%
$, they coincide with the case that was discussed by Yuen in \cite{YuenCNSNS}.

\begin{proof}
[Proof of Theorem \ref{thm1}]According to the Lemma, it is easy to check that
the function (\ref{e10}) satisfies the continuity equation (\ref{h2}). In the
following, we shall validate that the function (\ref{e10}) also holds for the
momentum equation (\ref{h3}).

For the $i$-th momentum equation of the Navier-Stokes equations (\ref{h3}), by
defining an elliptically symmetric variable via
\begin{equation}
s=\underset{k=1}{\overset{N}{\sum}}\frac{(x_{k}-d_{k})^{2}}{a_{k}^{2}(t)}.
\end{equation}
Now we proceed with $\gamma=2$, then we obtain
\begin{align}
&  \rho\left[  \frac{\partial u_{i}}{\partial t}+\sum_{k=1}^{N}u_{k}%
\frac{\partial u_{i}}{\partial x_{k}}\right]  -\kappa_{1}\frac{\partial
}{\partial x_{i}}\left(  \rho^{\theta}\nabla\cdot\vec{u}\right)  -\kappa
_{2}\nabla\cdot\left(  \rho^{\theta}\nabla u_{i}\right)  +\kappa\frac
{\partial}{\partial x_{i}}\rho^{\theta}\\[0.1in]
&  =\rho\left\{  \frac{\partial}{\partial t}\left[  \frac{\dot{a}_{i}}{a_{i}%
}(x_{i}-d_{i})+\dot{d}_{i}\right]  +\left[  \frac{\dot{a}_{i}}{a_{i}}%
(x_{i}-d_{i})+\dot{d}_{i}\right]  \frac{\partial}{\partial x_{i}}\left[
\frac{\dot{a}_{i}}{a_{i}}(x_{i}-d_{i})+\dot{d}_{i}\right]  \right\}
\nonumber\\[0.1in]
&  \quad\quad-\kappa_{1}\theta\sum_{k=1}^{N}\frac{\dot{a}_{k}}{a_{k}}%
\rho^{\theta-1}\frac{\partial\rho}{\partial x_{i}}-\kappa_{2}\theta
\rho^{\theta-1}\frac{\dot{a}_{i}}{a_{i}}\frac{\partial\rho}{\partial x_{i}%
}+\kappa\theta\rho^{\theta-1}\frac{\partial\rho}{\partial x_{i}}%
\nonumber\\[0.1in]
&  =\rho\left\{  \left[  \left(  \frac{\ddot{a}_{i}}{a_{i}}-\frac{\dot{a}%
_{i}^{2}}{a_{i}^{2}}\right)  (x_{i}-d_{i})+\ddot{d}_{i}+\frac{\dot{a}_{i}^{2}%
}{a_{i}^{2}}(x_{i}-d_{i})\right]  -\theta\rho^{\theta-2}\left(  \kappa_{1}%
\sum_{k=1}^{N}\frac{\dot{a}_{k}}{a_{k}}+\kappa_{2}\frac{\dot{a}_{i}}{a_{i}%
}-\kappa\right)  \frac{\partial\rho}{\partial x_{i}}\right\}
\nonumber\\[0.1in]
&  =\rho\left\{  \left[  \left(  \frac{\ddot{a}_{i}}{a_{i}}-\frac{\dot{a}%
_{i}^{2}}{a_{i}^{2}}\right)  (x_{i}-d_{i})+\ddot{d}_{i}+\frac{\dot{a}_{i}^{2}%
}{a_{i}^{2}}(x_{i}-d_{i})\right]  -\theta\left(  \kappa_{1}\sum_{k=1}^{N}%
\frac{\dot{a}_{k}}{a_{k}}+\kappa_{2}\frac{\dot{a}_{i}}{a_{i}}-\kappa\right)
\frac{f(s)^{\theta-2}}{\left(  \underset{k=1}{\overset{N}{\Pi}}a_{k}\right)
^{\theta-2}}\frac{\partial}{\partial x_{i}}\frac{f(s)}{\underset{k=1}%
{\overset{N}{\Pi}}a_{k}}\right\} \nonumber\\[0.1in]
&  =\rho\left\{  \frac{\ddot{a}_{i}}{a_{i}}(x_{i}-d_{i})+\ddot{d}_{i}%
-2\theta\left(  \kappa_{1}\sum_{k=1}^{N}\frac{\dot{a}_{k}}{a_{k}}+\kappa
_{2}\frac{\dot{a}_{i}}{a_{i}}-\kappa\right)  \frac{f(s)^{\theta-2}\dot{f}%
(s)}{\left(  \underset{k=1}{\overset{N}{\Pi}}a_{k}\right)  ^{\theta-1}}\left(
\frac{x_{i}-d_{i}}{a_{i}^{2}}\right)  \right\} \nonumber\\[0.1in]
&  =\frac{\rho(x_{i}-d_{i})}{a_{i}^{2}}\left\{  \ddot{a}_{i}a_{i}%
-\frac{2\theta f(s)^{\theta-2}\dot{f}\left(  s\right)  }{\left(
\underset{k=1}{\overset{N}{\Pi}}a_{k}\right)  ^{\theta-1}}\left(  \kappa
_{1}\sum_{k=1}^{N}\frac{\dot{a}_{k}}{a_{k}}+\kappa_{2}\frac{\dot{a}_{i}}%
{a_{i}}-\kappa\right)  \right\}  +\rho\ddot{d}_{i}\nonumber\\[0.1in]
&  =\frac{\rho(x_{i}-d_{i})}{a_{i}^{2}}\left\{  \xi+2\theta f(s)^{\theta
-2}\dot{f}\left(  s\right)  \right\}  +\rho\ddot{d}_{i}%
\end{align}
with the $N$-dimensional generalized Emden dynamical system given by :
\begin{equation}
\left\{
\begin{array}
[c]{l}%
\ddot{a}_{i}(t)=\frac{-\xi\left[  \kappa_{1}\sum_{k=1}^{N}\frac{\dot{a}%
_{k}(t)}{a_{k}(t)}+\kappa_{2}\frac{\dot{a}_{i}(t)}{a_{i}(t)}-\kappa\right]
}{a_{i}(t)\left(  \underset{k=1}{\overset{N}{\Pi}}a_{k}(t)\right)  ^{\theta
-1}},\text{ for }i=1,2,\cdots,N\\
a_{i}(0)=a_{i0}>0,\text{ }\dot{a}_{i}(0)=a_{i1}%
\end{array}
\right.  \label{EmdenEmdenChaos}%
\end{equation}
with arbitrary constants $\xi$, $a_{i0}$ and $a_{i1}.$ \newline If we require
the function $f(s)$ satisfies the following differential equation:
\begin{equation}
\left\{
\begin{array}
[c]{l}%
\frac{\xi}{2\theta}+f(s)^{\theta-2}\dot{f}\left(  s\right)  =0\\
f(0)=\alpha\geq0,
\end{array}
\right.  \quad\quad\quad\text{ or }\quad\quad\quad\rho=0 \label{firstODEChaos}%
\end{equation}
then we can have
\begin{equation}
f(s)=\left\{
\begin{array}
[c]{c}%
\alpha e^{-\frac{\xi}{2\theta}s}\text{ }%
\ \ \ \ \ \ \ \ \ \ \ \ \ \ \ \ \ \ \ \ \ \ \ \ \ \ \ \ \ \ \text{for }%
\theta=1\\
\max\left(  \left(  -\frac{\xi(\theta-1)}{2\theta}s+\alpha\right)  ^{\frac
{1}{\theta-1}},\text{ }0\right)  \text{ for }\theta\neq1.
\end{array}
\right.
\end{equation}
Therefore, the function (\ref{e10}) is the drifting solution with elliptic
symmetry of the compressible Navier-Stokes equations with dependent-density viscosity.
\end{proof}

\section{Conclusion and Discussion}

Due to the importance of the Naiver-Stokes equations in various branches of
physics, many experts have paid great attention to them, especially to the
constructions of analytical solutions. For example, when the viscosity
coefficients are chosen by $h(\rho)=\kappa_{1}\rho^{\theta}, g(\rho)=0$, Yuen
obtained the self-similar solutions in \cite{YuenJMP08}. Yuen also constructed
some self-similar solutions with elliptic symmetry when the viscosity
variables are $h(\rho)=\mu$ and $g(\rho)=0$ in \cite{YuenCNSNS}. Guo and Xin
derived some analytical solutions when $h(\rho)=\rho^{\theta}$ and
$g(\rho)=(\theta-1)\rho^{\theta}$ in \cite{GuoX}. Interestingly, here we
successfully derived some drifting solutions with elliptic symmetry for the
compressible Navier-Stokes equations with general forms of viscosity
coefficients. Numerical simulations show that the analytical solutions
obtained can be applied to explain the drifting phenomena of propagations of
wave like Tsunamis in oceans. In addition, we would like to point out that the
velocity components $a_{i}$ are governed by the generalized Emden dynamical
system, which is quite different from that the classical Emden equations
obtained in \cite{YuenJMP08,YuenCNSNS}. Does the generalized Emden system have
any nice properties as the classical one? What is the relation between the
generalized Emden system and the Ermakov system? These problems will be deeply
considered in our future work.

\textbf{Acknowledgements:} This work is supported by the Fundamental Research
Funds KJ2013036 for the Central Universities, the Foundation LXYQ201201112 of
Nanjing Agricultural University and Research Grant MIT/SRG14/12-13 of the Hong
Kong Institute of Education.

\end{document}